\documentclass[a4paper]{jpconf}
\usepackage[latin1]{inputenc}
\usepackage{graphicx}
\usepackage{amssymb}

\newcommand{\eV}{\mathrm{~eV}}
\newcommand{\keV}{\mathrm{~keV}}
\newcommand{\meV}{\mathrm{~meV}}
\newcommand{\MeV}{\mathrm{~MeV}}
\newcommand{\GeV}{\mathrm{~GeV}}
\newcommand{\cL}{\mathcal{L}}
\newcommand{\dm}{\partial_\mu}
\newcommand{\mbb}{m_{\beta\beta}}

\begin{document}

\title{$\nu$MSM and its experimental tests}

\author{F Bezrukov}
\address{Institut de Théorie des Phénomènes Physiques,\\
  École Polytechnique Fédérale de Lausanne,
  CH-1015 Lausanne, Switzerland}
\ead{fedor@ms2.inr.ac.ru}

\begin{abstract}
  $\nu$MSM is a minimal renormalizable extension of the Standard Model
  by right handed neutrinos.  This model explains the neutrino
  oscillations and provides a candidate for the Dark Matter and a
  mechanism of baryon number generation in the Early Universe.  We
  discuss here existing constraints on the model and possible
  consequences for astrophysical and laboratory experiments.
\end{abstract}

\paragraph{Introduction.}

Extension of the Standard Model (SM) is inevitable.  Current evidences
include neutrino
oscillations
and existence of large amount of the Dark Matter in the
Universe.

Here we shall describe a minimal extension of the SM which explains
these effects.  This extension is not addressing the hierarchy problem
or other naturalness issues.  Let us note, that aside from a certain
amount of fine tuning, the SM is logically consistent
\cite{Shaposhnikov:2007nj}, so it is really needed only to explain
these new experimental facts.

The model we adopt was introduced in \cite{Asaka:2005an,Asaka:2005pn}.
It adds the singlet leptons to the SM with the mass scale below the
electroweak scale ($O(1)\keV-O(1)\GeV$), so one may expect that
experimental investigation is viable.  We will discuss possible
astrophysical and laboratory experimental checks of the model.

\paragraph{The $\nu$MSM model and its predictions.}

The $\nu$MSM adds three singlet right-handed neutrinos $N_I$ to the
SM in the most general renormalizable way
\begin{displaymath}
  \cL_{\nu\mathrm{MSM}} = \cL_{\mathrm{SM}}
    + \overline{N}_I i\dm\gamma^\mu N_I
    - F_{\alpha I} \overline{L}_\alpha N_I\overline{\Phi}
    - \frac{M_{I}}{2}\overline{N^c_I} N_I
    + \mathrm{h.c.}
  \;,
\end{displaymath}
where $\overline{\Phi}=\epsilon^{ab}\Phi^*_b$ is the Higgs doublet,
$L_\alpha$ are the left-handed lepton doublets ($\alpha=e,\mu,\tau$).
New coupling constants present in this Lagrangian are the Yukawa
couplings $F_{\alpha I}$ (giving rise to the Dirac masses for the
neutrinos $M^D=F\langle\Phi\rangle$) and the Majorana masses $M_I$.
This gives 18 new parameters, which are 3 Majorana masses, 3 Dirac
masses, 6 mixing angles and 6 CP-violating phases.

\begin{figure}
  \begin{center}
    \begin{minipage}[b]{0.49\textwidth}
      \includegraphics[width=\textwidth]{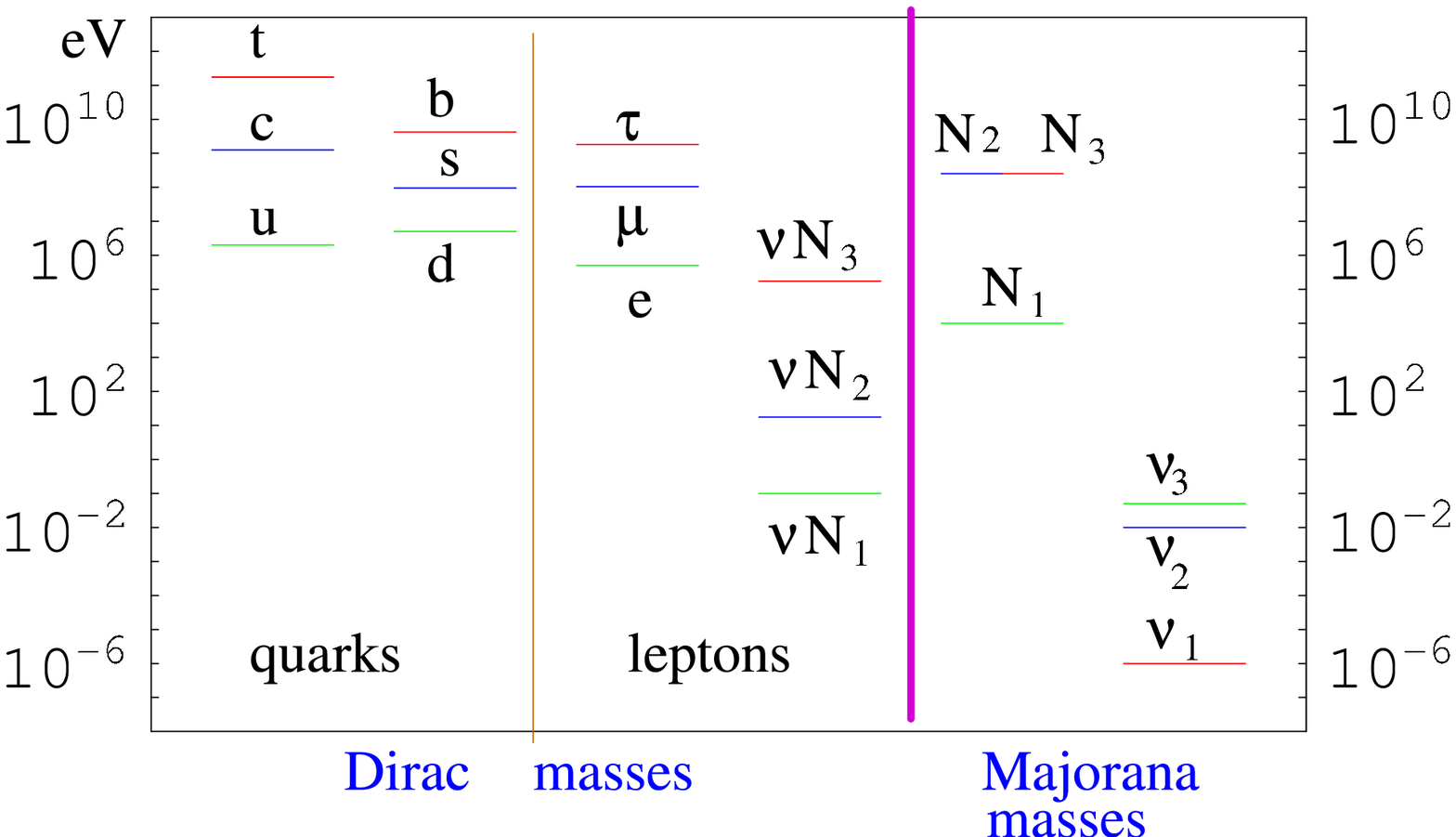}\\
      \caption{Typical mass scales of the $\nu$MSM\@.  The experimental
        values are given for the quark and the charged leptons.\\ \\}
      \label{fig:spectrum}
    \end{minipage}\hfill
    \begin{minipage}[b]{0.49\textwidth}
      \includegraphics[width=\textwidth]{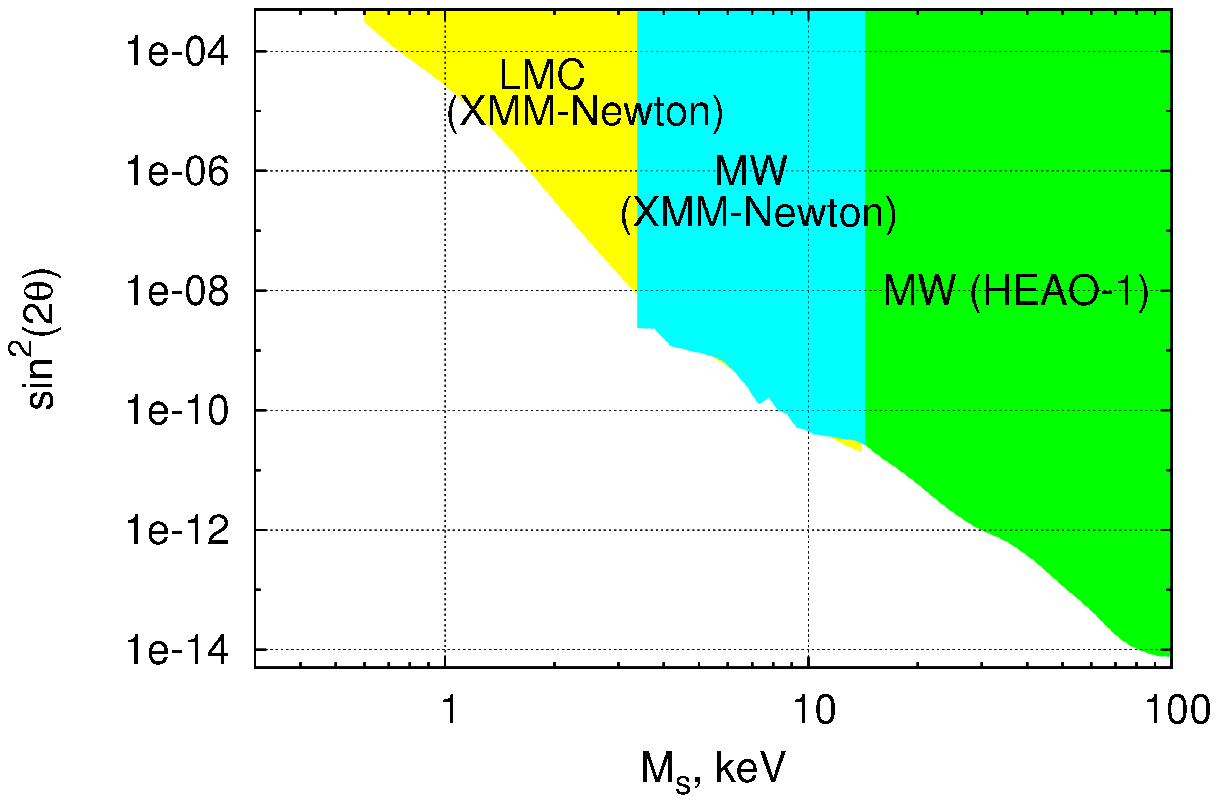}
      \caption{Upper bound on the active-sterile neutrino mixing angle form X-ray observations
        of the Large Magellanic Cloud (LMC) and the Milky Way (MW) by
        XMM-Newton and HEAO-1 satellites.}
      \label{fig:xray}
    \end{minipage}
  \end{center}
\end{figure}

This Lagrangian is the same that is used in the ordinary seesaw models
to explain small neutrino masses
\cite{Minkowski:1977sc,Yanagida:1980xy} and leads to the active
neutrino mass matrix of the form ${m^\nu=\left(M^D\right)^TM_I^{-1}M^D}$
and active-sterile neutrino mixing angles
$\theta=(M^D)^\dagger{}M_I^{-1}\ll1$.  However, it is not supposed
that the Majorana mass scale coincides with the grand unification
scale \cite{Shaposhnikov:2007nj}.  The typical mass spectrum of
the $\nu$MSM is given in the Fig.~\ref{fig:spectrum}.  It is
important to note, that while any mixing angles and mass splittings
observed in the active neutrino sector can be reproduced, the
cosmological requirements together with astrophysical constraints and
observed neutrino oscillation pattern fixes the absolute scale of the
active neutrino masses, predicting the hierarchical spectrum.

\paragraph{Dark matter and astrophysical constraints.}

The sterile neutrino $N_1$ with the mass $M_1<1\MeV$ decays mainly
into three active neutrinos ($2\nu\bar\nu$ or $2\bar\nu\nu$) by Z
boson exchange, with the life time
\(
  \tau_{N_1} =
    5\times10^{26}\mathrm{s}\left(\frac{1\keV}{M_1}\right)^5
    \left(\frac{10^{-8}}{\theta^2}\right)
\).  For mixing angle $\theta^2\simeq10^{-8}$
and $M_1\simeq 1\div10\keV$ we get the life time much larger than the
age of the Universe.  Such neutrino may play a role of the Warm Dark
Matter (WDM) particle.

The mass of the WDM particle is natural to choose of the order of
$O(10)\keV$.  This range is favourable to solve the problem of missing
satellite galaxies and cuspy profiles in the Cold Dark Matter
scenarios
\cite{Bode:2000gq,Gilmore:2007fy,Goerdt:2006rw,Moore:1999nt}.  Low
mass is also favoured by the thermal production mechanism via
active-sterile oscillations \cite{Dodelson:1993je}.  However, there is
really no upper bound if the DM neutrino mass is produced by other
mechanisms \cite{Shi:1998km,Shaposhnikov:2006xi,BGSHinfl}.  There is
an astrophysical lower bound of 0.3~keV
\cite{Lin:1983vq,Dalcanton:2000hn,Tremaine:1979we} and a stronger (but
model dependent) bound from Lyman-$\alpha$ observations
$M_1>10\div14\keV\left(\frac{\langle p_s\rangle}{\langle
    p_a\rangle}\right)$, where $\langle p_s\rangle$ and $\langle
p_a\rangle$ are the average momenta of the sterile and active
neutrinos \cite{Seljak:2006qw,Hansen:2001zv,viel:063534,Viel:2006kd}.

Another major constraint \cite{Boyarsky:2006fg} is provided by the
observation of the photons from the radiative decay $N_1\to\nu\gamma$
with the width \(
\Gamma(N_1\to\nu+\gamma)=1.38\times10^{-22}\sin^2(2\theta)
\left(\frac{M_1}{1\keV}\right)^5 \mathrm{s}^{-1} \;.  \) Thus, the
dark matter halos should emit the gamma rays with the energy
$E=M_1/2$.  The resulting constraint is presented in
Fig.~\ref{fig:xray}.
This constraint, immediately leads to the conclusion that the lightest
\emph{active} neutrino mass is extremely small, about $10^{-5}\eV$ or
smaller.

At present X-ray observations provide the strongest constraint on the
$\nu$MSM\@.  The discussion of the sensitivity of the future Space
missions can be found in \cite{Boyarsky:2006hr}.

\paragraph{Neutrinoless double beta decay.}

As far as the $\nu$MSM provides neutrinos with Majorana masses, the
neutrinoless double beta decay is possible.  The contribution of the
DM neutrino $N_1$ to the rate of the decay is negligible
\cite{Bezrukov:2005mx}, so the prediction for the effective Majorana
mass $\mbb$ for neutrinoless double beta decay coincides with the
usual 3 neutrino analysis \cite{Pascoli:2005zb} with strictly
hierarchical spectrum, i.e.\ $1.3\meV<\mbb^{NH}<3.4\meV$ in normal
neutrino mass hierarchy and $13\meV<\mbb^{IH}<50\meV$ in inverted
hierarchy.  In particular, this means that discovery of the double
beta decay at the rate corresponding to $\mbb>50\meV$ would definitely
contradict the $\nu$MSM model.

One should note, however, that $\mbb<13\meV$ or $<1.3\meV$ can be
explained in the framework of some modifications of the $\nu$MSM, that
is the model with large entropy release \cite{Asaka:2006ek} or with
relatively light $N_{1,2}\lesssim100\MeV$ \cite{Bezrukov:inprep}.

\paragraph{Kinematic study of beta decays.}

Previous two methods provide only indirect clues for the $\nu$MSM\@.
It is hard to overestimate the importance of a direct experimental
evidence for the sterile neutrinos.  However, experiments with
creation and subsequent detection of the DM neutrino in $\nu$MSM are
rendered impossible by extremely low allowed values of the mixing
angle, see Fig.~\ref{fig:xray}.  The only possibility left for
laboratory experiments is detailed kinematic study of the processes
creating $N_1$.  Measuring the momenta of all initial and final
particles in nuclear beta decay, except neutrino, allows to determine
the neutrino mass in each single event \cite{Bezrukov:2006cy}.
Experimental techniques based on time of flight measurements that
allow to determine recoil energy of the ion are being currently used
in experiments in atomic physics \cite{Ullrich1997,Dorner2000}.  In
such experiments the precision of momenta measurements reaches
0.2~keV, which is sufficient to distinguish between sterile and active
neutrinos.

Detailed study of such experiments is needed, which addresses the
problem of density of the source of the cold atoms (extremely high
statistics is needed) and the background from bremsstrahlung emission
of keV photons.

\begin{figure}
  \begin{center}
    \includegraphics[width=0.333\textwidth]{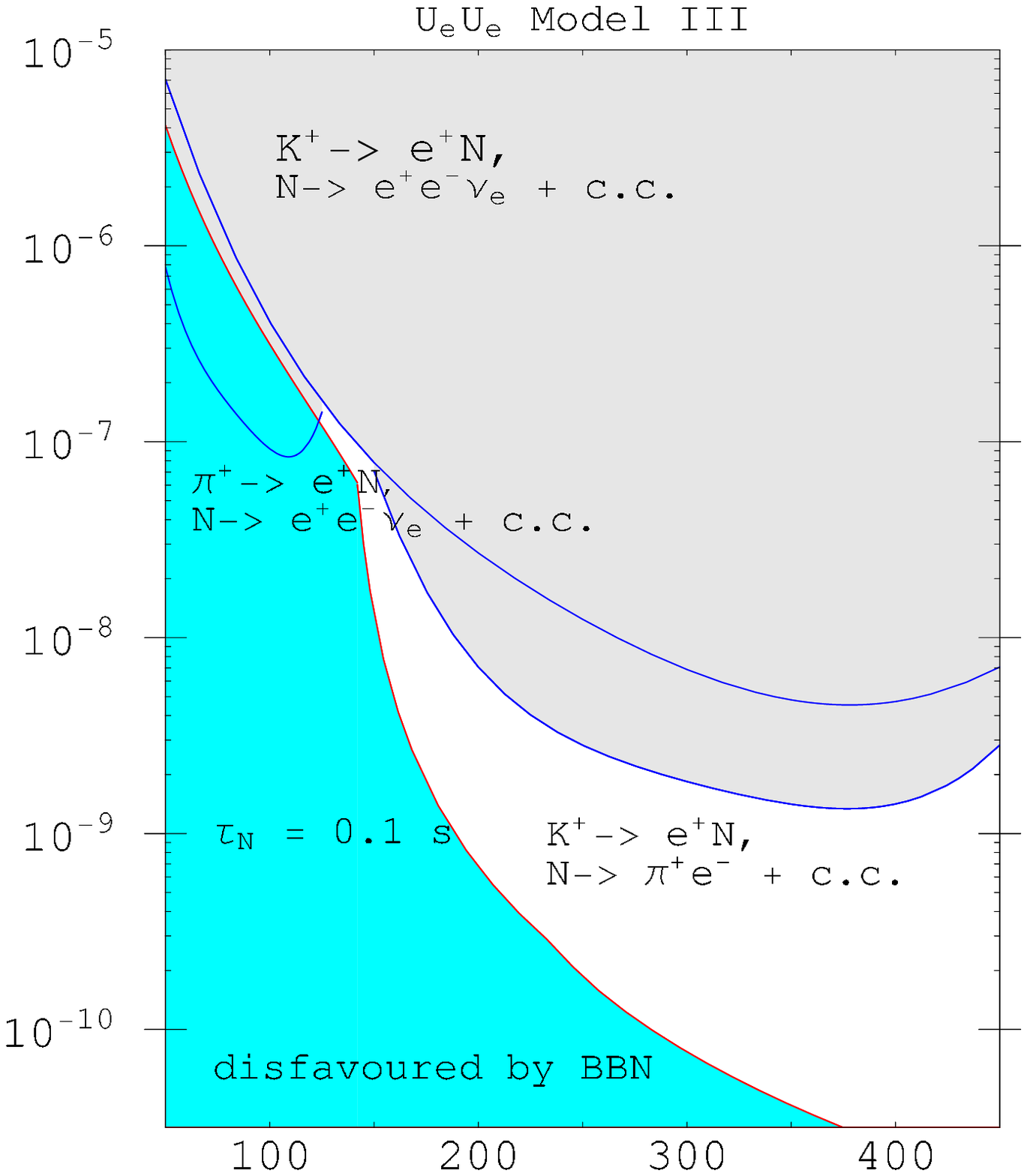}%
    \includegraphics[width=0.333\textwidth]{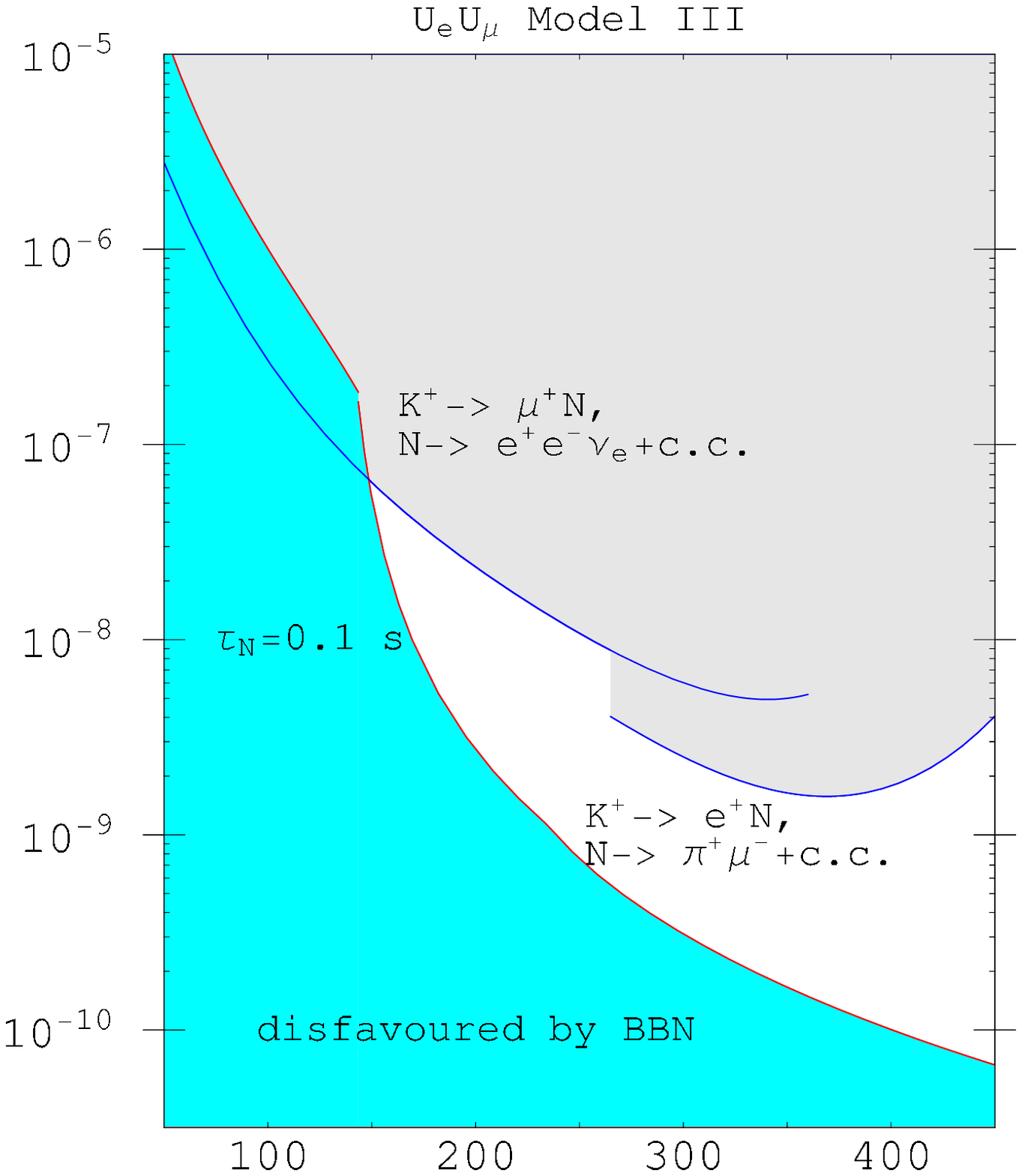}%
    \includegraphics[width=0.333\textwidth]{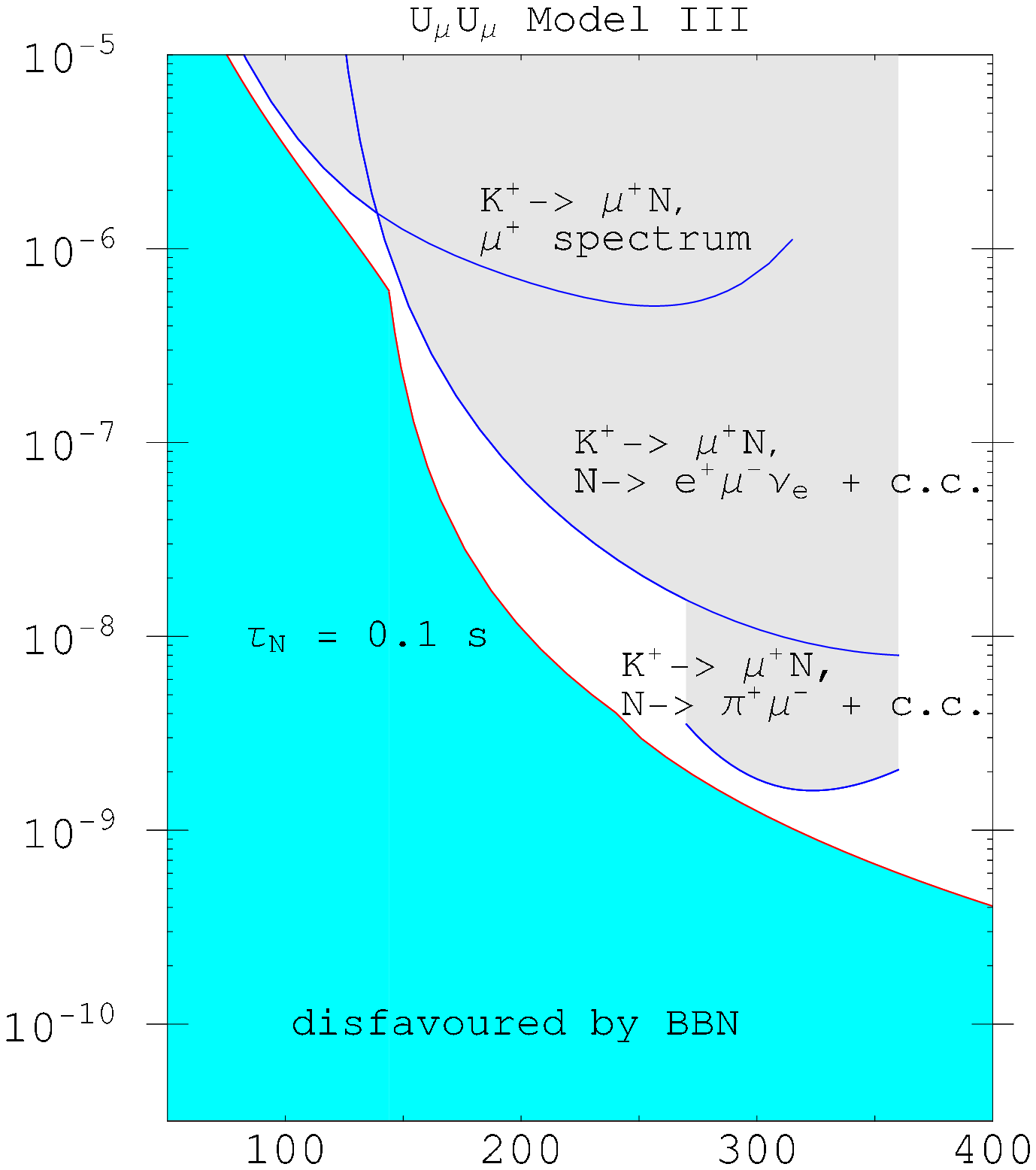}
    \caption{Limits on the mixing angles $U_\alpha={M^D_\alpha}^2/M_{2,3}$
      for the sterile neutrinos $N_{2,3}$ from the BBN and direct
      experimental search.  The currently allowed region is white
      \cite{Gorbunov:2007ak}.}
    \label{fig:ps191}
  \end{center}
\end{figure}

\paragraph{Heavy sterile neutrinos in the long base line experiments.}

Other two sterile neutrinos $N_{1,2}$ can be used to generate the
baryon asymmetry of the universe \cite{Asaka:2005pn}.  It proceeds via
resonant generation of the lepton number in the sterile neutrino
sector which is then transferred to active neutrino sector and
converted to baryon asymmetry via B+L violating electroweak sphaleron
transitions.

This mechanism requires that $N_{1,2}$ are very degenerate in mass and
that their Yukawa couplings are small enough so that they do not
thermalize before freeze-out of the sphaleron transitions (otherwise
the generated asymmetry is washed out).  Certain amount of CP
violation in the mixing matrix of the sterile neutrinos is also
required, but it seems to have no sizable observational effect for
laboratory experiments.

The mixing angles for $N_{1,2}$ have also the \emph{lower} bound from
the requirement that they decay before the Big Bang nucleosynthesis
(BBN), otherwise they would change the abundances of light elements in
the Universe.  This bound is quite close to the direct experimental
limits on sterile neutrinos with masses below the kaon mass.  And
sterile neutrinos $N_{1,2}$ with masses below the pion mass are
already excluded.  The experimental signature in such experiments is
appearance of $e^+e^-$ or $\mu^+\mu^-$ pairs from the decay of the
heavy sterile neutrino which is created in the decay of a charged kaon
in a beam dump.  Existing limits from CERN PS191 experiment are given
in Fig.~\ref{fig:ps191}.  Further searches look promising and can be
performed together with long baseline neutrino oscillation experiments
\cite{Gorbunov:2007ak}.

As a summary, the $\nu$MSM model provides explanation to neutrino
oscillations, DM and baryon asymmetry of the Universe.  At the same
time it has nontrivial experimental consequences, which can be checked
in future astrophysical and laboratory accelerator experiments.

\ack

Author is grateful to M. Shaposhnikov and D. Gorbunov for numerous
valuable discussions.

\section*{References}
\bibliographystyle{iopart-num}
\bibliography{all,inprep}

\end{document}